\documentclass[letterpaper,preprint,notoc]{JHEP} 
\usepackage{axodraw}

\newcommand\fverb{\setbox\pippobox=\hbox\bgroup\verb}
\newcommand\fverbdo{\egroup\medskip\noindent%
			\fbox{\unhbox\pippobox}\ }
\newcommand\fverbit{\egroup\item[\fbox{\unhbox\pippobox}]}
\newbox\pippobox

\newskip\humongous \humongous=0pt plus 1000pt minus 1000pt

\newif\ifdtup

\newcounter{eqnumber}[section]

\def\qb{{\bar{q}}}

\def\Tr{\, {\rm tr}}

\def\ksl{\not{\hbox{\kern-2.3pt $k$}}}
\def\lsim{\buildrel < \over {_\sim}}

\def\e{\epsilon}
\def\musq{\mu^2}

\def\eps{\epsilon}

\def\Ord{{\cal O}}
\def\cm{{\cal M}}

\def\Nf{{N_{\! f}}}
\def\la{\langle}
\def\ra{\rangle}
\def\RS{{\scriptscriptstyle\rm R\!.S\!.}}

\def\bom#1{{\mbox{\boldmath $#1$}}}
\def\MSbar{\overline{\rm MS}}
\def\lr{\leftrightarrow}
\def\li#1{{\mathop{\rm Li}\nolimits}_#1}
\def\Li{\mathop{\rm Li}\nolimits}
\def\ggtogg{gg \to \gamma\gamma}
\def\Sublead{{\rm \scriptscriptstyle SL}}
\def\Lead{{\rm \scriptscriptstyle L}}
\def\alphas{\alpha_s}

\def\eqn#1{eq.~(\ref{#1})}

\def\rg{r_\Gamma}

\def\Boxfour{{\rm Box}^{(4)}}
\def\Boxsix{{\rm Box}^{(6)}}
\def\Boxeight{{\rm Box}^{(8)}}

\def\Trifour{{\rm Tri}^{(4)}}
\def\Trisix{{\rm Tri}^{(6)}}

\def\Bubfour{{\rm Bub}^{(4)}}
\def\Bubsix{{\rm Bub}^{(6)}}

\def\spa#1.#2{\left\langle#1\,#2\right\rangle}
\def\spb#1.#2{\left[#1\,#2\right]}
\def\lor#1.#2{\left(#1\,#2\right)}

\preprint{
  hep-ph/0109078 \\
  SLAC--PUB--8976\\
    UCLA/01/TEP/16\\
    September, 2001}

\title{Two-Loop Amplitudes for Gluon Fusion into Two Photons}

\author{Z. Bern,\thanks{Research supported by the US Department of 
Energy under grant DE-FG03-91ER40662.} 
\ A. De Freitas$^*$ \\
Department of Physics and Astronomy \\
UCLA, Los Angeles, CA 90095-1547}

\author{L. Dixon\thanks{Research supported by the US Department of 
Energy under grant DE-AC03-76SF00515.}
\\
Stanford Linear Accelerator Center\\
Stanford University\\
Stanford, CA 94309}

\abstract{
We present the two-loop matrix elements for the scattering of two
gluons into two photons in QCD.  These matrix elements will enter into
improved estimates of the QCD background to Higgs production at the
LHC when the Higgs decays into two photons. The photon mode is
especially important if $M_H < 140$ GeV.  We also give the amplitudes
for the crossed process, $g\gamma \to g\gamma$.}


\dedicated{\centerline{Submitted to JHEP}}

\begin{document}

\section{Introduction}
\label{IntroSection}

One of the most pressing problems in particle physics today is to
determine the nature of electroweak symmetry breaking.  Experiments in
the coming decade at the Fermilab Tevatron and the CERN Large Hadron
Collider (LHC) will address this problem, in particular by searching
for one or more Higgs bosons.  There are good reasons to think that at
least one such particle will be fairly light.  In the Standard Model
the Higgs boson is constrained to be light by its influence on
precision electroweak measurements, $m_H \lsim 205$--$230$ GeV at 95\%
CL~\cite{HiggsRadCorr}.  In the Minimal Supersymmetric Standard Model,
the lightest Higgs boson is predicted to have a mass below about 135
GeV~\cite{SusyHiggs}; over much of the parameter space it has
properties reasonably similar to the Standard Model Higgs boson.
Finally, there are hints from LEP at $2.9 \sigma$ significance of a
Higgs boson at $m_H = 115$ GeV~\cite{LEPHiggs}.

To aid in the search for the Higgs boson, it is useful to have a 
detailed understanding of the Standard Model backgrounds.  
At the LHC the most important mode for discovering a Higgs boson with 
$m_H < 140$ GeV is via its decay into two 
photons~\cite{Higgsgammagamma,HiggsBkgdgammagamma}.
The irreducible two-photon background due to QCD, $pp \to \gamma\gamma X$, 
will be determined experimentally at the LHC, so
the Higgs search can proceed even with imprecise theoretical knowledge of it.
Nevertheless, it is still of interest to have robust theoretical
predictions prior to the experiments to help optimize Higgs search
strategies.  In this paper we provide the two-loop matrix elements for
$gg \to \gamma\gamma$ with massless quarks in the loop,
which form a central ingredient in an improved prediction of the 
irreducible background.

The process $pp \to \gamma\gamma X$ proceeds at lowest order via the
partonic subprocess $q\qb \to \gamma\gamma$, which is independent of
$\alphas$.  The next-to-leading-order (NLO) corrections to this subprocess
have been incorporated into a number of Monte Carlo
programs~\cite{TwoPhotonBkgd1}.  However, at the LHC, due to the high
gluon luminosity, formally higher order corrections involving gluon
initial states are sizable.  The contribution from the lowest order $gg
\to \gamma\gamma$ amplitudes (from one-loop box diagrams), though of order
$\alphas^2$, is very comparable in size to the $q\qb \to \gamma\gamma$
contribution~\cite{HiggsBkgdgammagamma,AmetllerDicusWillenbrock,
TwoPhotonBkgd1}.  Thus a calculation of the $gg \to \gamma\gamma$
subprocess at {\it its} next-to-leading-order (which is formally N$^3$LO
as far as the whole process $pp \to \gamma\gamma X$ is concerned), should
lead to a significant reduction in the uncertainty on the total cross
section.  A calculation of the background, incorporating the two-loop
matrix elements presented in this paper, will appear in a forthcoming
publication~\cite{HiggsBkgdPaper}.  In the range of di-photon invariant
masses relevant for the Higgs search, 90--150 GeV, quark masses may be
neglected.  The $u$, $d$, $s$, $c$ and $b$ quark masses are all much less
than the scale of the process, while the top quark contribution is
negligible until the invariant mass approaches $2m_t \approx 350$ GeV.

The subprocess $g\gamma \to g\gamma$ does not possess quite the same
phenomenological significance as $gg \to \gamma\gamma$.   However, it
does receive a large power-enhanced correction at two loops in the 
forward scattering limit, due the exchange of two gluons, which represents
the beginning of the Reggeization of this amplitude.  The
amplitudes for $gg \to \gamma\gamma$ and $g\gamma \to g\gamma$
are of course related by crossing symmetry, but writing the
full two-loop amplitude in a crossing-symmetric form would be more
cumbersome, so we shall present each case separately.  

Calculating two-loop four-point amplitudes involving more than a
single kinematic variable is a relatively new art.  The first
calculations of this type in gauge theory were for the special cases of 
gluon-gluon scattering with maximal
supersymmetry~\cite{BRY} and maximal helicity violation in
QCD~\cite{AllPlusTwo}.  More recently, more general calculations of
interferences of two-loop amplitudes with tree amplitudes in QED and
in QCD have appeared.  In QED the interferences for $e^+ e^-
\to \mu^+ \mu^-$ and $e^+ e^- \to e^+
e^-$~\cite{BhabhaTwoLoop} have been computed. In a tour de force
calculation, Anastasiou, Glover, Oleari, and Tejeda-Yeomans have
provided the interferences for all QCD $2\to 2$ parton
processes~\cite{GOTY2to2,GOTYgggg}.

The two-loop Feynman diagrams for the $g g \to \gamma \gamma$
matrix elements presented in this paper are similar to those required
for gluon-gluon scattering~\cite{GOTYgggg}, $gg \to gg$, except that
many of the non-Abelian diagrams are not present.  In the $gg \to
\gamma \gamma$ case, the tree amplitudes vanish and the one-loop
amplitudes give the leading order contributions.  Thus the {\it
next}-to-leading order contributions to this process require a
different interference, of two-loop amplitudes with one-loop
amplitudes.  Instead of evaluating this interference directly, we have
computed the two-loop $\ggtogg$ amplitudes in a helicity basis.  We
employed a unitarity- or cut-based
technique~\cite{CutBased,BRY,AllPlusTwo} to generate the required loop
momentum integrals.  These integrals were then evaluated using
recently developed techniques.

Two important technical breakthroughs which have provided the required
integrals for general $2 \to 2$ scattering in the massless
case are the calculations of the dimensionally regularized scalar
double box integrals with planar~\cite{PBScalar} and
non-planar~\cite{NPBScalar} topologies and all external legs massless,
and the development of reduction algorithms for the same types of
integrals with loop momenta in the numerator (tensor
integrals)~\cite{PBReduction,NPBReduction,GRReduction}.  Related
integrals which also arise in the reduction procedure have been
computed in refs.~\cite{IntegralsAGO}.  Taken together, these results
are sufficient to compute all loop integrals required for 
$2\to 2$ massless scattering amplitudes at two loops. 

A helpful development for performing explicit two-loop calculations is a
general formula due to Catani for the infrared divergence appearing in
any two-loop QCD amplitude~\cite{Catani}.  By appropriately adjusting
group theory factors, it is straightforward to apply Catani's QCD
formula to amplitudes including external photons.  Because of the vanishing 
of the tree amplitudes, the infrared divergences for $gg \to \gamma\gamma$
and $g\gamma \to g\gamma$ are much tamer than those of a typical
QCD process at two loops.  Indeed, Catani's formula collapses
to a form previously derived for one loop amplitudes~\cite{GieleGlover,KST}.  
We present the amplitudes in terms of Catani's formula for the infrared 
divergences, plus finite remainders for each independent helicity
configuration.

In the physical next-to-leading-order correction
to the $gg\to\gamma\gamma$ subprocess of $pp\to\gamma\gamma X$,
the infrared divergences from the two-loop amplitude cancel against 
those arising from phase-space integration of
the (square of) the one-loop amplitudes for 
$gg \to \gamma\gamma g$~\cite{FiveGluon,ggGamGamg},
after factorizing the initial-state collinear singularities.
Both these ``virtual'' and ``real'' divergences are effectively the same 
as encountered at one loop, so standard one-loop 
formalisms~\cite{OneLoopIR} can be employed to obtain the
cross section~\cite{HiggsBkgdPaper}.

This paper is organized as follows.  In section~\ref{IRSection} we
review the infrared structure of the two-loop $gg \to \gamma \gamma$ and
$g \gamma \to g \gamma$ amplitudes.
The corresponding one-loop amplitudes, which appear in the formula for 
the infrared divergences of the two-loop amplitudes, are presented in
section~\ref{OneLoopSection}.  In sections~\ref{TwoLoopgggamgamSection}
and~\ref{TwoLoopggamggamSection}
we present analytic results for the finite remainders of the two-loop 
$gg \to \gamma \gamma$ and $g \gamma \to g \gamma$ helicity amplitudes,
respectively.  
In section~\ref{ChecksSection} we discuss the checks we performed on our
results.  Finally, in section~\ref{ConclusionsSection} we present our
conclusions.

\section{Review of infrared structure}
\label{IRSection}

In presenting the amplitudes it is convenient to separate the infrared
divergent parts from the finite parts.  Dimensionally regulated
two-loop amplitudes for four massless particles in $D=4-2\e$ dimensions
generically contain poles in $\e$ up to $1/\e^4$.  Catani has presented 
a general formula for the structure of infrared divergences of any QCD
amplitude~\cite{Catani}.  With minor adjustments, Catani's formula is
also valid for the case of mixed amplitudes of QED and QCD.  We shall
therefore adopt his notation in presenting our results.

We work with ultraviolet renormalized amplitudes, and employ the
$\MSbar$ running coupling for QCD, $\alphas(\mu^2)$.  Since the tree
amplitudes for the process under consideration vanish, for purposes
of renormalization we only need the one-loop relation between the bare
coupling $\alphas^u$ and renormalized coupling $\alphas(\mu^2)$,
\begin{equation}
\alphas^u \, \mu_0^{2\e} \, S_\e \ =\ 
\alphas(\mu^2) \, \mu^{2\e} \Biggl[ 1 - \alphas(\mu^2) { \beta_0 \over \e}
   + \Ord(\alphas^2(\mu^2)) \Biggr] \,,
\label{OneLoopCoupling}
\end{equation}
where $S_\e = \exp[\e (\ln4\pi + \psi(1))]$ and 
$\gamma = -\psi(1) = 0.5772\ldots$ is Euler's constant.
The coefficient appearing in the QCD beta function is 
\begin{equation}
  \beta_0 = {11 C_A - 4 T_R \Nf \over 12 \pi} \,, \qquad
\label{QEDBetaCoeffs}
\end{equation}
where $\Nf$ is the number of light (massless) quarks,
$C_A = N$ for $SU(N)$ and $T_R = 1/2$ for fundamental representation 
Dirac fermions.

The renormalized QCD corrections to the $gg \to \gamma \gamma$ 
amplitude discussed in this
paper may be expanded as
\begin{eqnarray}
\cm_{gg \to \gamma \gamma}(\alphas(\mu^2), \alpha, \mu^2;\{p\}) 
&=&  4\pi\alpha \, 
\Biggl[
 { \alphas(\mu^2) \over 2\pi } 
\cm_{gg \to \gamma \gamma}^{(1)}(\mu^2;\{p\}) \label{RenExpand}\\
&& \null  \hskip1cm
+ \biggl( { \alphas(\mu^2) \over 2\pi } \biggr)^2
 \cm_{gg \to \gamma \gamma}^{(2)}(\mu^2;\{p\})
+ \Ord(\alphas^3(\mu^2))
\Biggr] \,. \nonumber
\end{eqnarray}
where $ \cm_{gg \to \gamma \gamma}^{(L)}(\mu^2;\{p\}) $ 
is the $L$th loop contribution.
Since there is no direct coupling between gluons and photons, the
expansion starts at one loop.  

The QCD $\MSbar$ counterterm which is subtracted from the bare
two-loop amplitude to obtain the renormalized  
$\cm_{gg \to \gamma \gamma}^{(2)}(\mu^2;\{p\})$ is ({\it cf.}
\eqn{OneLoopCoupling})
\begin{equation}
 \hbox{C.T.} = 
      \, { 11 N - 2 \Nf \over 6 } \, 
   {1\over\eps}  \cm_{gg \to \gamma \gamma}^{(1)}(\mu^2;\{p\}) \,. 
\label{CT1}
\end{equation}
The relative simplicity of the ultraviolet subtraction term is due to
the vanishing of the tree-level amplitudes.

\FIGURE{
\begin{picture}(102,80)(0,0)
\Text(10,5)[r]{1} \Text(10,74)[r]{2} 
\Text(94,81)[l]{3} \Text(87,5)[l]{4} 
\Line(29,22)(64,22) \Line(29,57)(64,57)
\ArrowLine(29,22)(29,57) \Line(64,22)(64,57)
\Gluon(29,22)(12,5){2.5}{3} \Gluon(12,74)(29,57){2.5}{3}
\Photon(81,5)(64,22){2.5}{3}
\CArc(76,69)(5,0,90) \ArrowArc(76,69)(5,90,180) \CArc(76,69)(5,180,360)
\Photon(64,57)(72.4,65.4){2.5}{2}
\Photon(79.6,72.6)(88,81){2.5}{2}
%
\end{picture}
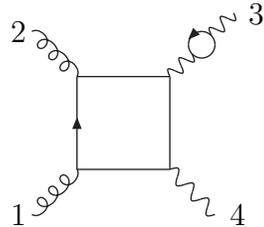
\caption{These diagrams give QED divergences.  They are exactly canceled by 
conventional on-shell renormalization.} \label{QEDDivergences}}

There is also a QED divergence associated with the diagrams in
Figure~\ref{QEDDivergences}.  In dimensional regularization with massless
fermions, these diagrams would vanish by virtue of containing a scale
free integral. This vanishing represents a cancellation between
infrared and ultraviolet divergences.  In QCD with $\MSbar$
renormalization such diagrams (with the bubble on a gluon leg)
are thus taken to vanish.  However, if one
renormalizes QED in the conventional on-shell scheme, to avoid the
infrared divergences one should retain the fermion masses in the
external bubbles.  Now the bubble integral is nonzero and ultraviolet
divergent.  But this divergence, and indeed the entire integral, is
exactly canceled by the on-shell-scheme counterterm, precisely
because the external leg is a real, on-shell photon.  In the on-shell
scheme, the coupling constant should of course be set to $\alpha
\equiv \alpha(0) = 1/137.036\ldots$.  This value should then be used for
all the QED couplings associated with real, external photons.

The infrared divergences of a renormalized two-loop amplitude, 
for the case where the tree amplitude vanishes, are~\cite{Catani},
\begin{equation}
| \cm_n^{(2)}(\mu^2; \{p\}) \ra_{\RS} = 
{\bom I}^{(1)}(\e, \mu^2; \{p\}) 
\; | \cm_n^{(1)}(\mu^2; \{p\}) \ra_{\RS} 
+ | \cm_n^{(2){\rm fin}}(\mu^2; \{p\}) \ra_{\RS} \,,
\label{TwoloopCatani}
\end{equation}
where $|\cm_n^{(L)}(\mu^2; \{p\}) \ra_{\RS}$ is a color space
vector representing the renormalized $L$ loop amplitude.   
The subscript $\RS$ stands for the choice of renormalization scheme,
and $\mu$ is the renormalization scale. These color space vectors give the
amplitudes via,
\begin{equation}
{\cal M}_n(1^{a_1},\dots,n^{a_n}) \equiv
\la a_1,\dots,a_n \,
| \, \cm_n(p_1,\ldots,p_n)\ra \,,
\label{MnVec}
\end{equation}
where the $a_i$ are color indices.  The divergences of ${\cal M}_n$
are encoded in the color operators 
${\bom I}^{(1)}(\e,\mu^2;\{p\})$.

In pure QCD, the operator ${\bom I}^{(1)}(\e,\mu^2;\{p\})$ is given by
\begin{equation}
{\bom I}^{(1)}(\e,\mu^2;\{p\}) =  \frac{1}{2}
{e^{-\e \psi(1)} \over \Gamma(1-\e)} \sum_{i=1}^n
\, \sum_{j \neq i}^n \, 
{\bom T}_i \cdot {\bom T}_j \biggl[ {1 \over \e^2}
 + {\gamma_i \over {\bom T}_i^2 } \, {1 \over \e} \biggr] 
\biggl( \frac{\mu^2 e^{-i\lambda_{ij} \pi}}{2 p_i\cdot p_j} \biggr)^{\e}
 \,,
\label{CataniGeneral}
\end{equation}
where $\lambda_{ij}=+1$ if $i$ and $j$ are both incoming or outgoing
partons and $\lambda_{ij}=0$ otherwise. The color charge ${\bom T}_i =
\{T^a_i\}$ is a vector with respect to the generator label $a$, and an
$SU(N)$ matrix with respect to the color indices of the outgoing 
parton $i$.  For external gluons $T^a_{cb} = i f^{cab}$,
so ${\bom T}_i^2 = C_A = N$, and
\begin{equation}
\gamma_g = {11\over 6}\, C_A - {2 \over 3} T_R \, \Nf \,. 
\label{QCDValues}
\end{equation}

The two processes we consider here are
\begin{eqnarray}
 g(-p_1,-\lambda_1)\ +\ g(-p_2,-\lambda_2)\ 
&\to&\ \gamma(p_3,\lambda_3)\ +\ \gamma(p_4,\lambda_4), 
\label{gggamgamlabel} \\
 g(-p_1,-\lambda_1)\ +\ \gamma(-p_2,-\lambda_2)\ 
&\to&\ g(p_3,\lambda_3)\ +\ \gamma(p_4,\lambda_4), 
\label{ggamggamlabel}
\end{eqnarray}
using an ``all-outgoing'' convention for the momentum ($p_i$)
and helicity ($\lambda_i$) labeling.  
The Mandelstam variables are $s = (p_1+p_2)^2$, 
$t = (p_1+p_4)^2$, and $u = (p_1+p_3)^2 $.  
To apply \eqn{CataniGeneral} to these processes, one
may convert two of the gluon legs to photons by setting 
${\bom T}_\gamma \cdot {\bom T}_i \to 0$,\  
${\bom T}_i \cdot {\bom T}_\gamma \to 0$
to obtain the simplified formula
\begin{equation}
{\bom I}^{(1)}_{gg \to \gamma\gamma} (\e,\mu^2;\{p\}) =
-  N {e^{-\e \psi(1)} \over \Gamma(1-\e)} 
\biggl[ {1 \over \e^2} 
     +  \biggl( {11\over 6} 
              - {1\over 3} {N_f\over N} \biggr) {1 \over \e}  \biggr]
\biggl({\mu^2\over -s} \biggr)^{\e}  \,.
\label{Catani1}
\end{equation}
The color structure of the two-gluon, two-photon amplitude can only
be proportional to $\delta^{a_1 a_2}$.  Hence 
${\bom I}^{(1)}_{gg \to \gamma\gamma}$ is necessarily
proportional to the identity operator, and the color space 
language is actually unnecessary.  In the $s$-channel
where $s>0$ one should use the usual analytic continuation
\begin{equation}
(-s-i \varepsilon)^{-\eps} =  |s|^{-\eps} e^{i \pi \eps\, \Theta(s)} \,.
\label{SimpleContinuation}
\end{equation}
For the crossed process $g\gamma \to g\gamma$, the same formulae apply
with the obvious modification,
\begin{equation}
{\bom I}^{(1)}_{g\gamma \to g\gamma} (\e,\mu^2;\{p\}) =
-  N {e^{-\e \psi(1)} \over \Gamma(1-\e)} 
\biggl[ {1 \over \e^2} 
     +  \biggl( {11\over 6} 
              - {1\over 3} {N_f\over N} \biggr) {1 \over \e}  \biggr]
\biggl({\mu^2\over -u} \biggr)^{\e}  \,.
\label{Catani1ggamggam}
\end{equation}

The renormalized two-loop amplitudes may thus be separated into 
divergent parts (which also absorb some finite terms) and a finite remainder,
\begin{eqnarray}
{\cal M}_{ {gg \to \gamma\gamma}}^{(2)} 
       &=& {\bom I}^{(1)}_{gg \to \gamma\gamma}
        (\e,\mu^2;\{p\}) {\cal M}_{{gg \to \gamma\gamma}}^{(1)} 
  + {\cal M}_{ {gg \to \gamma\gamma}}^{(2)\rm fin} \,,
\label{FiniteInfinite} \\
{\cal M}_{ {g\gamma \to g\gamma}}^{(2)} 
       &=& {\bom I}^{(1)}_{g\gamma \to g\gamma}
        (\e,\mu^2;\{p\}) {\cal M}_{{g\gamma \to g\gamma}}^{(1)} 
  + {\cal M}_{ {g\gamma \to g\gamma}}^{(2)\rm fin} \,.
\label{FiniteInfiniteggamggam}
\end{eqnarray}
The finite remainders 
${\cal M}_{{gg \to \gamma\gamma}}^{(2)\rm fin}$ and
${\cal M}_{ {g\gamma \to g\gamma}}^{(2)\rm fin}$ 
will be presented in sections~\ref{TwoLoopgggamgamSection} 
and~\ref{TwoLoopggamggamSection}.
Since ${\bom I}^{(1)}$ contains $1/\e^2$ poles, the above decomposition
requires the expansion of ${\cal M}^{(1)}$ through $\Ord(\e^2)$,
which will be given in section~\ref{OneLoopSection}.

We present the two-loop amplitudes for definite external helicities, 
using 't~Hooft-Veltman (HV) dimensional regularization~\cite{HV}.  At
one loop, scheme conversions between the various flavors of
dimensional regularization have been extensively
discussed~\cite{SchemeConvert}.  Following a
similar strategy, one may convert our results in the 't~Hooft-Veltman
scheme to the conventional dimensional regularization (CDR) scheme.
In the CDR scheme one would compute the interference of say
${\cal M}_{ {gg \to \gamma\gamma}}^{(2)}$ with 
${\cal M}_{ {gg \to \gamma\gamma}}^{(1)}$, summed over all colors and
helicities.
This CDR interference may be obtained from our HV result by using 
\eqn{FiniteInfinite} for ${\cal M}_{ {gg \to \gamma\gamma}}^{(2)}$
and replacing the HV interference 
${\cal M}_{ {gg \to \gamma\gamma}}^{(1)} \times
{\cal M}_{ {gg \to \gamma\gamma}}^{(1)\, *}$
by the same quantity evaluated in the CDR scheme, while leaving
${\bom I}^{(1)}$ and the finite remainder 
${\cal M}_{ {gg \to \gamma\gamma}}^{(2)\rm fin}$ 
of section~\ref{TwoLoopgggamgamSection} unaltered.
The same considerations apply for $g\gamma\to g\gamma$ as well.
When constructing the cross section for 
$pp \to \gamma \gamma X$, the difference between the HV
and the CDR scheme amplitudes is cancelled by a similar scheme dependence
in the real emission terms, and at the end only the finite
remainder contributes from the two-loop amplitudes~\cite{HiggsBkgdPaper}.  


\section{One-loop amplitudes}
\label{OneLoopSection}

As noted above, the decompositions~(\ref{FiniteInfinite}) 
and~(\ref{FiniteInfiniteggamggam})
require the series expansions of the one-loop $gg \to \gamma \gamma$
and $g\gamma \to g \gamma$ amplitudes through $\Ord(\e^2)$,
A simple way to obtain these expansions is to express them in terms of
four-gluon amplitudes which are known to all orders in
$\e$~\cite{BernMorgan}, in terms of integral functions whose series
expansions are known to the appropriate 
order~\cite{AllPlusTwo,BhabhaTwoLoop}.  The fermion loop contributions 
to $gg \to gg$ satisfy the color decomposition
\begin{eqnarray}
{\cal M}_{gg \to gg}^{(1)f}(1,2,3,4) &=& 
 \Nf \, \sum_{\sigma\in S_3}  
\Tr[T^{a_{\sigma(1)}} T^{a_{\sigma(2)}} T^{a_{\sigma(3)}} T^{a_{\sigma(4)}}]
    \nonumber \\
&& \null \hskip 1 cm \times 
A^{(1)}(\sigma(1), \sigma(2), \sigma(3), \sigma(4))\,,   
\label{FourGluonColor}
\end{eqnarray}
where $\Nf$ is the number of quark flavors and the sum of
permutations $\sigma$ runs over the six non-cyclic permutations of
external legs.  We have adjusted the normalization factors in the
amplitudes to match the conventions of ref.~\cite{Catani}.

One may obtain amplitudes with photons from \eqn{FourGluonColor} simply by
replacing the appropriate color matrices with the identity matrix 
and altering the normalizations.  In particular, the amplitudes where
any two legs are gluons and the other two legs are photons are all the 
same,
\begin{eqnarray}
{\cal M}_{gg \to \gamma \gamma}^{(1)}  
= {\cal M}_{g\gamma \to g\gamma}^{(1)}
= 2 \, \delta^{a_1 a_2}\, 
\biggl(\sum_{i=1}^{\Nf} Q_i^2 \biggr) \, M^{(1)}\,,
\label{TwoPhotonColor}
\end{eqnarray}
where $Q_i$ are the electric charges ($2/3$ for up-type quarks and
$-1/3$ for down-type quarks) and
\begin{equation}
M^{(1)}
= \sum_{\sigma\in S_3} \, 
   A^{(1)}(\sigma(1), \sigma(2), \sigma(3), \sigma(4))\,.
\label{GluonToPhoton}
\end{equation}
The permutations $\sigma$ run over the same six orderings of external
legs as in \eqn{FourGluonColor}, while the overall factor of 2 arises
from our non-standard normalization of color matrices, requiring an
extra factor of $\sqrt{2}$ for each gluon converted to a photon.  In
the permutation sum the ultraviolet and infrared divergences appearing
in the four-gluon amplitudes cancel, leaving a finite expression.

The amplitudes are evaluated in the spinor helicity
formalism~\cite{SpinorHelicity}.  It is convenient to extract overall 
spinor phases from each helicity amplitude, 
\begin{eqnarray}
A^{(1)}(1^{\lambda_1}, 2^{\lambda_2}, 3^{\lambda_3}, 4^{\lambda_4}) 
& = & S_{\lambda_1 \lambda_2 \lambda_3 \lambda_4} \,
A^{(1)}_{\lambda_1 \lambda_2 \lambda_3 \lambda_4} \,, \nonumber  \\
M^{(1)}(1^{\lambda_1}, 2^{\lambda_2}, 3^{\lambda_3}, 4^{\lambda_4}) 
& = & S_{\lambda_1 \lambda_2 \lambda_3 \lambda_4} \,
M^{(1)}_{\lambda_1 \lambda_2 \lambda_3 \lambda_4} \,,
\label{SpinorPhaseExtract}
\end{eqnarray}
where the $\lambda_i$ signify the helicities of each leg and
\begin{eqnarray}
S_{++++} &=& i{\spb1.2 \spb3.4 \over \spa1.2 \spa3.4} \,,
  \label{SpinorPhases} \\
S_{-+++}  =  i{\spa1.2 \spa1.4 \spb2.4 \over \spa3.4 \spa2.3 \spa2.4} \,,
   \hskip0.3cm
S_{+-++} &=& i{\spa2.3 \spa2.4 \spb3.4 \over \spa1.4 \spa3.1 \spa3.4} \,,
   \hskip0.3cm
S_{++-+}  =  i{\spa3.2 \spa3.4 \spb2.4 \over \spa1.4 \spa2.1 \spa2.4} \,,
    \nonumber \\
S_{--++}  =  i{\spa1.2 \spb3.4 \over \spb1.2 \spa3.4} \,,
   \hskip1.25cm
S_{-+-+} &=& i{\spa1.3 \spb2.4 \over \spb1.3 \spa2.4} \,,
   \hskip1.25cm
S_{+--+}  =  i{\spa2.3 \spb1.4 \over \spb2.3 \spa1.4} \,. \nonumber
\end{eqnarray}
Our conventions here are that all external legs are
outgoing; for incoming legs one should reverse the helicities of those
legs. The spinor inner products~\cite{SpinorHelicity,MPReview} are
$\spa{i}.j = \langle i^- | j^+\rangle$ and $\spb{i}.j = \langle i^+|
j^-\rangle$, where $|i^{\pm}\rangle$ are massless Weyl spinors of
momentum $k_i$, labeled with the sign of the helicity.  They are
anti-symmetric, with norm $|\spa{i}.j| = |\spb{i}.j| = \sqrt{s_{ij}}$,
where $s_{ij} = 2k_i\cdot k_j$.  From these definitions 
it follows that the $S_{\lambda_1 \lambda_2 \lambda_3 \lambda_4}$ are
phases.

The fermion loop contributions to the one-loop four-gluon helicity amplitudes,
in a form valid to all orders in $\eps$ (with four-dimensional external 
momenta $p_i$) are~\cite{BernMorgan}
%
%
\begin{eqnarray}
A^{(1)}_{++++} & = &
    \eps (1-\eps) \,
     \Boxeight(s,t) 
\,,   \nonumber\\
  A^{(1)}_{-+++} & = &
     -  { t(u-s) \over s u } \eps \, \Trisix(s)
     - { s(u-t) \over t u } \eps \,  \Trisix(t) 
      \nonumber \\
&& \null 
      - {t-u \over s^2 } \eps \, \Bubsix(s)
      - {s-u \over t^2 } \eps \, \Bubsix(t)
      + {s t \over 2 u } \eps \, \Boxsix(s,t)  \nonumber \\
&& \null 
      + \eps (1-\eps) \Boxeight(s,t)
\, ,    \nonumber \\
  A^{(1)}_{++-+} & = &
  A^{(1)}_{-+++} 
\,,  \nonumber \\
  A^{(1)}_{--++} & = &
      - {1\over 2} s\, \eps \, \Boxsix(s,t) - {s \over 2 t}\Bubfour(t) 
      \nonumber \\
&& \null
  + 
  {s\over t^2}  \Bubsix(t)
  - {1 \over t} \, \eps \, \Bubsix(t) 
  +  \eps(1-\eps) \, \Boxeight(s,t)
\,,     \nonumber \\
  A^{(1)}_{-+-+} & = &
{u \over s t}\biggl[
   {1\over 2} t  \Bubfour(s)
  +  {1\over 2}s  \Bubfour(t)
  -  {1\over 2}s t (1-\eps) \, \Boxsix(s,t)
       \nonumber \\
&& \null 
      + {s t(s-t) \over u^2 } \eps \, \Trisix(t)
      + {s t(t-s) \over u^2 } \eps \, \Trisix(s)
      -{ t s^2 \over u^2 } \Bubfour(t)
         \nonumber \\
&& \null 
      -{ s t^2 \over u^2 } \Bubfour(s)
      -{ s \over t  }  \Bubsix(t)
      -{ t \over s  }  \Bubsix(s)
      +{ s \over u } \eps\, \Bubsix(t)
        \nonumber \\
&& \null 
      +{ t \over u } \eps\, \Bubsix(s)
      +{ s t \over u } \Trisix(t)
      +{ s t \over u } \Trisix(s)
        \nonumber \\
&& \null 
      +{ s^2 t^2 \over u^2 } \Boxsix(s,t)
      +{ s t \over u }\, \eps (1-\eps) \, \Boxeight(s,t)
  \biggr]
\label{OneLoopGlueAmplitudes}
\,. 
\end{eqnarray}
Here ${\rm Bub}^{(n)}(s)$, ${\rm Tri}^{(n)}(s)$ and 
${\rm Box}^{(n)}(s,t)$ are
the one-loop bubble, triangle and box scalar integrals, evaluated in
$D = n - 2\eps$ dimensions.  The remaining helicity configurations can
all be obtained using parity and relabelings.  In the CDR scheme one
would need additional $\eps$-helicities~\cite{KosowerHelicity}.

The bubble and box integrals that appear in the above amplitudes are 
\begin{eqnarray}
 \Bubfour(s)  &=& {\rg \over \eps (1-2 \eps)} (-s)^{-\eps} \, , \nonumber\\
 \Bubsix(s) &=& - {\rg \over 2\eps (1-2\eps) (3-2\eps)} 
       (-s)^{1-\eps} \, , \nonumber\\
\Trifour(s) &=& -{\rg \over \eps^2} \, (-s)^{-1-\eps}\,, \label{IntDefs}\\
 \Trisix(s) &=& -{\rg (-s)^{-\eps} \over 2 \eps (1- 2\eps)
                            (1 - \eps)} \,, \nonumber\\
\end{eqnarray}
where
\begin{eqnarray}
\rg & = & e^{-\eps \psi(1)} \,
 {\Gamma(1+\eps) \Gamma^2(1-\eps) \over \Gamma(1-2\eps)} 
  \nonumber \\
    & = &
     1 - {1\over2} \zeta_2 \, \e^2 - {7\over3} \zeta_3 \, \e^3
       - {47\over16} \zeta_4 \, \e^4 + \Ord(\e^5) \,,
\end{eqnarray}
with
\begin{equation}
\zeta_s \equiv \sum_{n=1}^\infty n^{-s} \,, \qquad \quad
\zeta_2 = {\pi^2\over6} \,, \qquad \zeta_3 = 1.202057\ldots, \qquad
\zeta_4 = {\pi^4\over90} \,,
\label{ZetaValues}
\end{equation}
and we have kept the full dependence on $\eps$ in the integrals.  In
the $s$-channel where $s>0$ the functions are given by the 
analytic continuation (\ref{SimpleContinuation}).

The box integrals in various dimensions appearing in
\eqn{OneLoopGlueAmplitudes} are related via a dimension-shifting
formula~\cite{DimShift} valid to all orders in $\eps$,
\begin{eqnarray}
\Boxsix(s,t) & =&  {1\over 2 \, (-1+2\eps) u} \, 
     \Bigl(s t \, \Boxfour(s,t) - 2 t\, \Trifour(t) 
             - 2 s \, \Trifour(s) \Bigr) \,, \nonumber \\
\Boxeight(s,t) & = &  {1\over 2 \, (-3+2\eps) u} \,
     \Bigl(s t \, \Boxsix(s,t) - 2 t\, \Trisix(t)
             - 2 s \, \Trisix(s) \Bigr) \,. 
\end{eqnarray}
Since the $D=6-2\e$ scalar box integral is completely finite as $\e\to0$, 
it is convenient to express the other box integrals in terms
of it. This isolates all divergences to triangle and bubble
integrals.  In the $u$-channel ($s<0$, $t<0$), where the functions are 
manifestly real, the expansion of the six-dimensional box through
$\Ord(\e^2)$ is~\cite{AllPlusTwo,BhabhaTwoLoop}
\begin{eqnarray}
\Boxsix(s,t) &= &
{ \rg u^{-1-\e} \over 2 (1-2\e) } 
       \Biggl[ 
   {1\over2} \Bigl( (V-W)^2 + \pi^2 \Bigr)
 \nonumber \\
&& \null 
 + 2 \e  \biggl( \li3(-v) - V \li2(-v) - {V^3\over3}  
                                         - {\pi^2\over2} V \biggr) 
 \nonumber \\
&& \null 
 - 2 \e^2 \biggl( \li4(-v) + W \li3(-v) - {1\over2} V^2 \li2(-v)
             - {1\over8} V^4 - {1\over6} V^3 W 
  \nonumber \\
&& \null 
+ {1\over4} V^2 W^2 
             - {\pi^2\over4} V^2 - {\pi^2\over3} V W - 2 \zeta_4 \biggr)
  + \hbox{$(s \lr t)$}\ \Biggr]\ +\ \Ord(\e^3), \hskip 2 cm 
\label{D6BoxEuclid}
\end{eqnarray}
where
\begin{equation}
 v = {s\over u} \, , \quad w = {t\over u} \, , \quad 
 V = \ln\biggl(-{s\over u}\biggr) \, , \quad 
 W = \ln\biggl(-{t\over u}\biggr) \, .
\label{uChannelvwVWdef}
\end{equation}
In the $s$-channel ($s>0$, $t<0$) an analytic continuation 
of the box integral yields,
\begin{eqnarray}
\Boxsix(s,t) &= &
{ \rg |s|^{-\e} \over u (1-2\e) } 
       \biggl\{ {1\over 2} X^2 
\nonumber \\ && \null
    + \e \biggr( - \li3(-x) +  X \li2(-x) -{1\over 3} X^3 +  \zeta_3
    + {1\over 2}  Y X^2 - {1\over 2} \pi^2 X \biggl) 
\nonumber \\ && \null
 - \e^2  \biggl(
         \li4(-x/y) 
        - \li4(-y)
        + \li3(-y) X
\nonumber \\ && \null
        + {1\over 2} \li2(-x) (X^2 + \pi^2)
        + {1\over 24} (Y^2 + \pi^2)^2 
       -  {1\over 6} Y^3 X 
       + {1\over 4} Y^2 X^2 
\nonumber \\ && \null
        + {1\over 3} X^3 Y
        - {1\over 8} (X^2 + \pi^2)^2
        + {\pi^2\over 3} X Y
        + {7\over 360} \pi^4 \biggr)
\nonumber \\ && \null
+ i \pi \biggl[ X + \e \Bigl( \li2(-x) +Y X - {1\over 2} X^2 
                - {\pi^2 \over 6} \Bigr)
\nonumber \\ && \null
    + \e^2 \Bigl(- \li3(-x) -  \li3(-y) - {1\over 2} Y X^2 
          + {1\over 6} X^3 
          +  \zeta_3 \Bigr) \biggr] \biggr\} \hskip 2 cm 
\nonumber \\ && \null 
 + \Ord(\e^3)\,, 
\label{D6Box}
\end{eqnarray}
where the variables appearing in the amplitudes are 
\begin{equation}
 x = {t\over s} \, , \quad y = {u\over s} \, , \quad 
 X = \ln\biggl(-{t\over s}\biggr) \, , \quad 
 Y = \ln\biggl(-{u\over s}\biggr) \, .
\label{VariableNames}
\end{equation}
In the permutation sum appearing in \eqn{TwoPhotonColor}, after
appropriate relabelings, both expansions of the box integrals
(\ref{D6BoxEuclid}) and (\ref{D6Box}) appear.

Through $\Ord(\eps^0)$ the amplitudes simply greatly and the
$M_{\lambda_1\lambda_2\lambda_3\lambda_4}^{(1)}$ reduce to, 
\begin{eqnarray}
M_{++++}^{(1)} &=&
   1 + \Ord(\e)
\,, \nonumber \\
M_{-+++}^{(1)} &=& M_{+-++}^{(1)} = M_{++-+}^{(1)} = M_{+++-}^{(1)} =
   1 + \Ord(\e),
 \nonumber \\
M_{--++}^{(1)} &=&
- {1\over2} {t^2+u^2\over s^2}
                  \Bigl[ \ln^2\Bigl({t\over u}\Bigr) + \pi^2 \Bigr]
		- {t-u\over s} \ln\Bigl({t\over u}\Bigr) - 1 + \Ord(\e)
\,,  \nonumber \\
M_{-+-+}^{(1)} &=&
 - {1\over2} {t^2+s^2\over u^2}
                  \ln^2\Bigl(-{t\over s}\Bigr)
		- {t-s\over u} \ln\Bigl(-{t\over s}\Bigr) - 1 \nonumber \\
&& \null \hskip 2 cm
	- i \pi \biggl[ {t^2+s^2\over u^2} \ln\Bigl(-{t\over s}\Bigr)
                      + {t-s\over u} \biggr] + \Ord(\e)
\,, \nonumber \\
M_{+--+}^{(1)}(s,t,u) &=&
M_{-+-+}^{(1)}(s,u,t)
\,.
\label{OneLoopFunctions}
\end{eqnarray}
%

\section{Finite parts of the two-loop $gg \to \gamma\gamma$ amplitudes}
\label{TwoLoopgggamgamSection}

A generic sample of two-loop Feynman diagrams for $gg \to \gamma
\gamma$ is shown in Figure~\ref{DiagramSampleFigure}.  We did not
evaluate the diagrams directly.  Instead we computed the unitarity
cuts in various channels, working to all orders in the dimensional
regularization parameter $\e = (4-D)/2$~\cite{CutBased}.  Essentially
we followed the approach first employed at two loops for the pure
gluon four-point amplitude with all identical
helicities~\cite{AllPlusTwo} and for $N=4$ supersymmetric
amplitudes~\cite{BRY}.  These amplitudes were simple enough that a
compact expression for the integrand could be given.  The fermion loop
contributions with all plus helicities are about as
simple~\cite{TwoLoopSUSY}.  However, for the generic helicity
configuration, the integrands become rather complicated. We therefore
used the general integral reduction algorithms developed for the
all-massless four-point
topologies~\cite{PBReduction,NPBReduction,GRReduction}, in order to
reduce the loop integrals to a minimal basis of master integrals.  To
incorporate polarization vectors of photons and gluons with definite
helicity requires some minor extensions of these 
techniques~\cite{ggggpaper}.

\FIGURE{
\begin{picture}(380,192)(0,0)
\Text(10,110)[r]{1} \Text(10,179)[r]{2} 
\Text(102,179)[l]{3} \Text(102,110)[l]{4} 
\Gluon(54,127)(29,127){2.5}{3} \Line(79,127)(54,127) 
\Gluon(29,162)(54,162){2.5}{3} \Line(54,162)(79,162)
\Gluon(29,127)(29,162){2.5}{4}
\Line(54,162)(54,127) \ArrowLine(79,127)(79,162)
\Gluon(29,127)(12,110){2.5}{3} \Gluon(12,179)(29,162){2.5}{3} 
\Photon(96,110)(79,127){2.5}{3} \Photon(79,162)(96,179){2.5}{3} 
%
\Text(152,110)[r]{1} \Text(152,179)[r]{2} 
\Text(244,179)[l]{3} \Text(244,110)[l]{4} 
\Line(196,127)(221,127) \Gluon(196,127)(171,127){2.5}{3} 
\Gluon(171,162)(196,162){2.5}{3} \ArrowLine(196,162)(221,162)
\Gluon(171,127)(171,162){2.5}{4} \Line(196,127)(221,162)
\Line(196,162)(207,146) \Line(210,143)(221,127)
\Gluon(171,127)(154,110){2.5}{3} \Gluon(154,179)(171,162){2.5}{3}
\Photon(238,110)(221,127){2.5}{3} \Photon(221,162)(238,179){2.5}{3} 
%
\Text(294,110)[r]{1} \Text(294,179)[r]{2} 
\Text(371,179)[l]{3} \Text(371,110)[l]{4} 
\Line(313,127)(348,127) \Line(313,162)(348,162)
\ArrowLine(313,127)(313,162) \Line(348,127)(348,162)
\Gluon(348,142)(326,162){2.5}{3}
\Gluon(313,127)(296,110){2.5}{3} \Gluon(296,179)(313,162){2.5}{3}
\Photon(365,110)(348,127){2.5}{3} \Photon(348,162)(365,179){2.5}{3} 
%
\Text(10,5)[r]{1} \Text(10,74)[r]{2}
\Text(102,74)[l]{3} \Text(102,5)[l]{4}
\ArrowLine(29,22)(29,57) \Line(79,22)(79,57)
\Line(29,22)(79,22) \Line(29,57)(44,57) \Line(64,57)(79,57) 
\GlueArc(54,57)(10,0,180){2.5}{4} \CArc(54,57)(10,180,0)
\Gluon(29,22)(12,5){2.5}{3} \Gluon(12,74)(29,57){2.5}{3}
\Photon(96,5)(79,22){2.5}{3} \Photon(79,57)(96,74){2.5}{3}
%
\Text(142,22.5)[r]{1} \Text(142,56.5)[r]{2}
\Text(254,74)[l]{3} \Text(254,5)[l]{4}
\Gluon(161,39.5)(186,39.5){2.5}{3}
\Gluon(186,39.5)(211,57){2.5}{3} \Gluon(186,39.5)(211,22){2.5}{3}
\Line(211,22)(211,57)
\Line(211,22)(231,22) \Line(211,57)(231,57) \ArrowLine(231,22)(231,57)
\Gluon(161,39.5)(145,22.5){2.5}{3} \Gluon(145,56.5)(161,39.5){2.5}{3}
\Photon(248,5)(231,22){2.5}{3} \Photon(231,57)(248,74){2.5}{3}
\Text(294,5)[r]{1} \Text(294,74)[r]{2} 
\Text(371,74)[l]{3} \Text(371,5)[l]{4} 
\Line(313,22)(348,22) \Line(335,57)(348,57)
\Line(313,22)(313,35) \ArrowLine(348,22)(348,57)
\Line(313,35)(335,57) \Gluon(313,57)(335,57){2}{3}
\Gluon(313,35)(313,57){2}{3}
\Gluon(313,22)(296,5){2.5}{3} \Gluon(296,74)(313,57){2.5}{3}                  
\Photon(365,5)(348,22){2.5}{3} \Photon(348,57)(365,74){2.5}{3}
%
\end{picture}

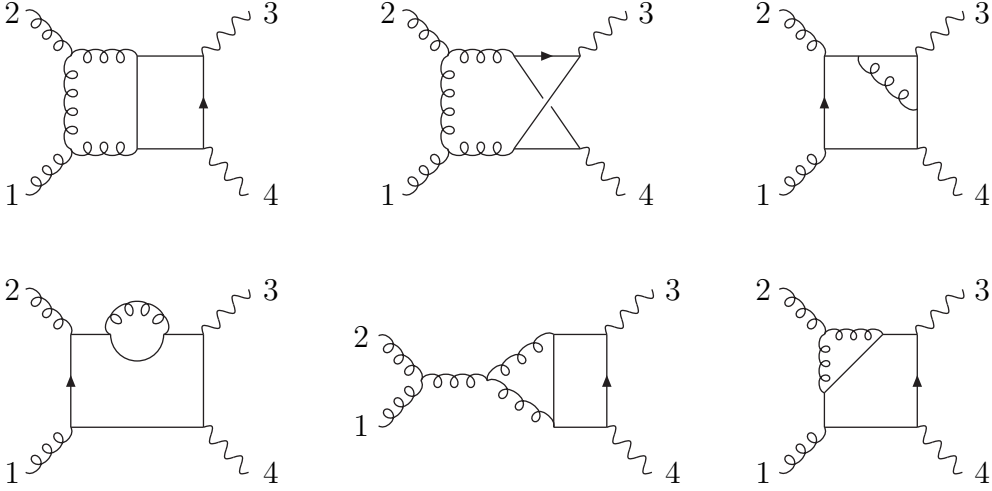
\caption{Some of the two-loop diagrams for $gg \to
\gamma \gamma$. The curly lines represent gluons while the
wavy ones photons.}\label{DiagramSampleFigure}}

We then expand the master integrals in a Laurent series in $\e$, which
begins at order $1/\e^4$. 
Many of the master integral expansions quoted in 
refs.~\cite{PBScalar,NPBScalar,PBReduction,NPBReduction,IntegralsAGO}
are in terms of Nielsen functions~\cite{NielsenRef}, 
\begin{equation}
S_{n,p}(x) = {(-1)^{n+p-1} \over (n-1)! \, p!} 
\int_0^1 {dt \over t} \ln^{n-1} t \, \ln^p(1-xt)\,,
\label{NielsenDef}
\end{equation}
with $n+p \leq 4$.  We have found it useful to express the results 
instead in terms of a minimal set of polylogarithms~\cite{Lewin},
\begin{eqnarray}
\Li_n(x) &=& \sum_{i=1}^\infty { x^i \over i^n }
          = \int_0^x {dt \over t} \Li_{n-1}(t)\,,  \\
\Li_2(x) &=& -\int_0^x {dt \over t} \ln(1-t) \,, 
\label{PolyLogDef}
\end{eqnarray}
with $n=2,3,4$, using relations such as~\cite{NielsenIds}
\begin{eqnarray}
S_{13}(x) &=& - \Li_4(1-x) + \ln(1-x) \Li_3(1-x) 
+ {1\over2} \ln^2(1-x) \Bigl( \Li_2(x) - \zeta_2 \Bigr) \nonumber\\
&& \null \hskip 1 cm 
+ {1\over3} \ln^3(1-x) \ln x + \zeta_4 \,, \nonumber\\
\null 
S_{22}(x) &=& \Li_4(x) - \Li_4(1-x) + \Li_4\biggl({-x\over1-x}\biggr)
- \ln(1-x) \Bigl( \Li_3(x) - \zeta_3 \Bigr) \nonumber\\
&& \null 
+ {1\over24} \ln^4(1-x) - {1\over6} \ln^3(1-x) \ln x
+ {1\over2} \zeta_2 \ln^2(1-x) + \zeta_4 \,, \nonumber\\
&& \null 
\hbox{for} \quad 0 < x < 1. 
\label{SomeNielsenRelations}
\end{eqnarray}

The analytic properties of the non-planar double box integrals
appearing in the amplitudes are somewhat
intricate~\cite{AllPlusTwo,NPBScalar}, since there is no Euclidean
region in any of the three kinematic channels for the $2 \to 2$
process.  We quote our results in the physical $s$-channel 
$(s > 0; \; t, \, u < 0)$ for the $gg\to\gamma\gamma$ 
kinematics~(\ref{gggamgamlabel}).

The dependence of the the finite remainder in \eqn{FiniteInfinite} on
quark charges, the renormalization scale $\mu$, $N$ and $\Nf$ 
may be extracted as,
\begin{eqnarray}
{\cal M}^{(2) \rm fin}_{gg \to \gamma\gamma} & = &
2 \delta^{ab} \, 
  \, \biggl( \sum_{j = 1}^\Nf Q_j^2 \biggr) \,
S_{\lambda_1 \lambda_2 \lambda_3 \lambda_4} \, \biggl[
   {11 N - 2 \Nf \over 6} \Bigl( \ln(\musq/s) + i \pi \Bigr) 
    M^{(1)}_{\lambda_1 \lambda_2 \lambda_3 \lambda_4}
\nonumber \\ && \null \hskip 4.5 cm 
  + N F^\Lead_{\lambda_1 \lambda_2 \lambda_3 \lambda_4} 
  - {1\over N} \, 
      F^\Sublead_{\lambda_1 \lambda_2 \lambda_3 \lambda_4} \biggr] \,,
\label{FiniteRemainder}
\end{eqnarray}
where the spinor phases are defined in \eqn{SpinorPhases} and coupling
constants have been extracted in \eqn{RenExpand}.  The $\mu$-dependence
in the first term in this expression is a consequence of 
renormalization group invariance.

The two-loop QED corrections to $gg \to \gamma \gamma$ 
require the same set of two-loop diagrams as the subleading-color 
QCD corrections.  In the QED case, external fermion bubble
insertions on the photon legs should be added.  However, as mentioned in 
section~\ref{OneLoopSection} these diagrams
are exactly cancelled by the ultraviolet counterterm in the usual
on-shell renormalization of QED (just as in the light-by-light
case~\cite{PhotonPaper}).  The renormalized amplitudes are free of 
infrared and ultraviolet divergences and are given by
\begin{equation}
{\cal M}^{(2) \rm \scriptscriptstyle QED}_{gg \to \gamma\gamma}  = 
4 \, \delta^{ab} \, 
    \, \biggl(\sum_{j = 1}^\Nf Q_j^4 \biggr) \,
S_{\lambda_1 \lambda_2 \lambda_3 \lambda_4} \, 
 F^\Sublead_{\lambda_1 \lambda_2 \lambda_3 \lambda_4} \,,
\label{QEDAmplitude}
\end{equation}
where a factor of $\alpha(\mu^2)$ replaces one factor of $\alphas(\mu^2)$
in the prefactor in front of ${\cal M}^{(2)}$ in \eqn{RenExpand}.  The
two-loop QCD and QED corrections to light-by-light
scattering via massless fermions are also proportional to
$F^\Sublead_{\lambda_1 \lambda_2 \lambda_3 \lambda_4}$~\cite{PhotonPaper}.  
(In ref.~\cite{PhotonPaper}
particles 1 and 2 are taken to be incoming, so the helicity labels for
legs 1 and 2 are reversed with respect to the labeling used here.)

The explicit forms for the $F^\Lead_{\lambda_1\lambda_2\lambda_3\lambda_4}$ 
appearing in \eqn{FiniteRemainder} are
\begin{equation}
 F^\Lead_{++++} =  
 {1\over 2} \,,  
\hskip 11.5 cm \null 
\label{FppppL}
\end{equation}
\begin{eqnarray}
F^\Lead_{-+++}  & = &  
     {1\over 8}  \biggl[ \biggl(2 + 4 {x\over y^2} - 5 {x^2\over y^2} \biggr) 
         ( (X + i \pi)^2 + \pi^2 )
       - (1 - x y)  ( (X-Y)^2 + \pi^2 ) \nonumber \\
&& \hskip 4 cm \null 
      + 2  \biggl( {9\over y} - 10 x \biggr)  (X + i \pi)  \biggr]
+ \Bigl\{t \leftrightarrow u \Bigr\} \,,
\hskip 2.2 cm \null 
\label{FmpppL}
\end{eqnarray}
\begin{eqnarray}
 F^\Lead_{++-+}  & = &  
     {1\over 8}  \biggl[ \biggl(2 + 6 {x\over y^2} - 3 {x^2\over y^2}\biggr) 
                ( (X + i \pi)^2 + \pi^2 )
        - (x-y)^2  ((X-Y)^2 + \pi^2) 
\nonumber \\ && \hskip 4 cm \null 
       + 2  \biggl({9\over y} - 8x \biggr)  (X + i \pi) \biggr]
+  \Bigl\{t \leftrightarrow u \Bigr\} \,, 
\nonumber
\hskip 2.2 cm \null 
\label{FppmpL} 
\end{eqnarray}
\begin{eqnarray}
 F^\Lead_{--++}  & = &  
 - (x^2+y^2) \biggl[ 4 \li4(-x) 
      + (Y - 3 X - 2 i\pi) \li3(-x)  
\nonumber \\ && \hskip2.0cm
      + ((X + i\pi)^2 + \pi^2) \li2(-x) 
      + {1\over 48} (X+Y)^4 
\nonumber \\ && \hskip2.0cm
      + i {\pi\over12} (X+Y)^3 
            + i {\pi^3\over2} X
      - {\pi^2\over12} X^2 - {109\over 720} \pi^4 \biggr] 
\nonumber \\ && \hskip-0.3cm
   + {1\over2} x (1 - 3 y) \biggl[
          \li3(-x/y) - (X-Y)  \li2(-x/y) - \zeta_3
       + {1\over 2}  Y  ((X-Y)^2 + \pi^2) \biggr] 
\nonumber \\ && \hskip-0.3cm
       + {1\over4} x^2 \Bigl[ (X-Y)^3 
                     + 3 (Y + i\pi) ( (X-Y)^2 + \pi^2 ) \Bigr] 
\nonumber \\ && \hskip-0.3cm
       + {1\over 8} \biggl( 14 (x-y) - {8\over y} + {9\over y^2} \biggr) 
                     ( (X + i\pi)^2 + \pi^2 ) 
\nonumber \\ && \hskip-0.3cm
       + {1\over 16}  ( 38 x y - 13 ) ( (X-Y)^2 + \pi^2 )
       - {\pi^2\over12}
       - {9\over4}  \biggl( {1\over y} + 2x \biggr) (X + i\pi) + {1\over 4}
\nonumber \\
&& \null
\hskip 4 cm +  \Bigl\{ t \leftrightarrow u \Bigr\} \,,
\label{FmmppL}
\end{eqnarray}
\begin{eqnarray}
 F^\Lead_{-+-+}  & = &  
   - 2 {x^2+1\over y^2} \biggl[  \li4(-x) - \zeta_4
       - {1\over 2} (X + i\pi) (\li3(-x) - \zeta_3 ) 
\nonumber \\ && \hskip2.0cm
       + {\pi^2\over6} \biggl( \li2(-x) - {\pi^2\over6} 
                             - {1\over2} X^2 \biggr)
       - {1\over 48} X^4 
\nonumber \\ && \hskip2.0cm
     + {1\over 24} (X + i\pi)^2 ( (X + i\pi)^2 + \pi^2 ) \biggr] 
\nonumber \\ && \hskip-0.3cm
   + 2 {3 (1-x)^2 - 2 \over y^2} \biggl[ 
            \li4(-x) + \li4(-x/y) - \li4(-y)
          - (Y + i\pi) ( \li3(-x) - \zeta_3 )  
\nonumber \\ && \hskip3.0cm
          + {\pi^2\over6} \biggl( \li2(-x) + {1\over2} Y^2 \biggr)
          - {1\over6} X Y^3 + {1\over24} Y^4 - {7\over360} \pi^4 \biggr]
\nonumber \\ && \hskip-0.3cm
    - {2\over 3} ( 8 - x + 30 {x\over y} ) \biggl[ 
            \li3(-y) - \zeta_3
         - (Y + i\pi) \biggl( \li2(-y) - {\pi^2\over6} \biggr)  
\nonumber \\ && \hskip3.1cm
         - {1\over2} X ((Y + i\pi)^2 + \pi^2) \biggr] 
\nonumber \\ && \hskip-0.3cm
    + {1\over 6} \biggl( 4 y + 27 + {42\over y} + {4\over y^2} \biggr)
                 \biggl[
           \li3(-x) - \zeta_3 
         - (X + i\pi) \biggl( \li2(-x) - {\pi^2\over6} \biggr) 
\nonumber \\ && \hskip4.5cm
         + i {\pi\over2} X^2 - \pi^2 X \biggr] 
\nonumber \\ && \hskip-0.3cm
  + {1\over 12} \biggl( 3 - {2\over y} - 12 {x\over y^2} \biggr) 
                (X + i\pi) ( (X + i\pi)^2 + \pi^2 ) 
\nonumber \\ && \hskip-0.3cm
  - {1\over 3} y (X + i\pi) ( (Y + i\pi)^2 + \pi^2 ) 
  + 2 \biggl( 1 + {2\over y} \biggr) 
           \Bigl( \zeta_3 - {\pi^2\over6} (Y + i\pi) \Bigr)
\nonumber \\ && \hskip-0.3cm
  + {1\over 24} \biggl(y^2 - 24 y + 44 - 8 {x^3\over y} \biggr) 
                        ( (X-Y)^2 + \pi^2 ) 
\nonumber \\ && \hskip-0.3cm
  - {1\over 24} \biggl(15 - 14 {x\over y} - 48 {x\over y^2} \biggr) 
                   ( (X + i\pi)^2 + \pi^2 ) 
\nonumber \\ && \hskip-0.3cm
  + {1\over 24} \biggl( 8 {x\over y} + 60 - 24 {y\over x} 
                      + 27 { y^2\over x^2} \biggr)
                   ( (Y + i\pi)^2 + \pi^2 )
  + {4\over 9} \pi^2 {x\over y}   
\nonumber \\ && \hskip-0.3cm
  + {1\over 12} ( 2 x^2 - 54 x - 27 y^2 ) 
       \biggl({1\over y} (X + i\pi) 
            + {1\over x} (Y + i\pi) \biggr) \,,
\label{FmpmpL}
\end{eqnarray}
where the last amplitude does not possess symmetry under $t \lr u$. 
In contrast to the one-loop case, the $-$$+$$+$$+$ case
is distinct from the $+$$+$$-$$+$ case, because at two loops 
external gluons can couple to a gluon internal to the diagram via
the non-abelian coupling, while photons cannot.

Similarly, the subleading color contributions in \eqn{FiniteRemainder}
are expressed in terms of the functions,
\begin{equation}
F^\Sublead_{++++} =  
- {3\over 2 }\,, 
 \hskip 11.5 cm \null 
\label{FppppSL}
\end{equation}
\begin{eqnarray}
F^\Sublead_{-+++}  & = &  
    {1\over 8}  \biggl[ { x^2 + 1\over y^2 } ((X + i \pi)^2 + \pi^2 )
             + {1\over 2} (x^2 + y^2) ((X-Y)^2 + \pi^2)  
\nonumber \\ && \hskip 4 cm \null
             - 4  \biggl({1\over y} - x \biggr)  (X + i \pi) \biggr] 
+ \Bigl\{t \leftrightarrow u \Bigr\} \,,
 \hskip 2.1 cm \null
\label{FmpppSL}
\end{eqnarray}
\begin{eqnarray}
 F^\Sublead_{++-+}  & = &  
 F^\Sublead_{+-++}\ =\ F^\Sublead_{+++-}\ =\ 
F^\Sublead_{-+++} \,, 
\hskip7.0cm
\label{FppmpSL}
\end{eqnarray}
\begin{eqnarray}
 F^\Sublead_{--++}  & = &  
 - 2 x^2 \biggl[ \li4(-x) + \li4(-y) 
  - (X + i\pi) \Bigl( \li3(-x) + \li3(-y) \Bigr)  
\nonumber \\ && \hskip1.2cm 
  + {1\over12} X^4 - {1\over3} X^3 Y + {\pi^2 \over 12} X Y 
  - {4 \over 90} \pi^4
  + i {\pi\over6} X \Bigl( X^2 - 3 X Y + \pi^2 \Bigr) \biggr] 
\nonumber \\ && \hskip-0.3cm
-(x-y) \Bigl( \li4(-x/y) - {\pi^2\over6} \li2(-x) \Bigr) 
\nonumber\\ && \hskip-0.3cm
 - x \biggl[ 2 \li3(-x) - \li3(-x/y) - 3 \zeta_3
      - 2 (X + i\pi) \li2(-x) 
\nonumber \\ && \hskip0.6cm 
      + (X-Y) ( \li2(-x/y) + X^2 ) 
      + {1\over 12} ( 5 (X-Y) + 18 i \pi) ((X-Y)^2 + \pi^2) 
\nonumber \\ && \hskip0.6cm 
      - {2\over 3}  X  (X^2 + \pi^2) - i \pi (Y^2 + \pi^2) \biggr]
\nonumber \\ && \hskip-0.3cm
  + { 1 - 2 x^2 \over 4 y^2 } ((X + i\pi)^2 + \pi^2)
  - {1\over 8} ( 2 x y + 3 ) ((X-Y)^2 + \pi^2)
  + { \pi^2 \over 12} 
\nonumber \\ && \hskip-0.3cm
  + \biggl( {1\over 2 y} + x \biggr) (X + i\pi) - {1\over 4} 
                    +  \Bigl\{ t \leftrightarrow u \Bigr\} \,,
\label{FmmppSL}
\end{eqnarray}
\begin{eqnarray}
 F^\Sublead_{-+-+}  & = &  
- 2 {x^2+1 \over y^2} \biggl[ 
       \li4(-x/y) - \li4(-y) 
    + {1\over2} (X - 2 Y - i\pi) ( \li3(-x) - \zeta_3 ) \nonumber \\
&& \hskip2.0cm
    + {1\over 24} ( X^4 + 2 i \pi X^3 - 4 X Y^3 + Y^4 
              + 2 \pi^2 Y^2 ) + {7\over 360} \pi^4 \biggr] \nonumber \\
&& \null
 - 2 {x-1\over y} \biggl[ \li4(-x) - \zeta_4
       - {1\over 2} (X + i\pi) (\li3(-x) - \zeta_3 )  \nonumber \\
&& \hskip2.0cm
       + {\pi^2\over 6} \Bigl( \li2(-x) - {\pi^2\over6} 
                              - {1\over2} X^2 \Bigr) 
       - {1\over 48} X^4 \biggr]  \nonumber \\
&& \null
 + \biggl(2 {x\over y} - 1\biggr) 
    \biggl[ \li3(-x) - (X + i\pi) \li2(-x) 
            + \zeta_3 - {1\over 6} X^3
            - {\pi^2\over 3} (X + Y) \biggr]  \nonumber \\
&& \null
 + 2 \biggl(2 {x\over y} + 1 \biggr) 
     \biggl[ \li3(-y) + (Y + i\pi) \li2(-x) - \zeta_3 
     + {1\over 4} X ( 2 Y^2 + \pi^2 )   \nonumber \\
&& \hskip2.8cm
          - {1\over 8} X^2 (X + 3 i\pi) \biggr]
     - {1\over 4} ( 2 x^2 - y^2 ) ((X-Y)^2 + \pi^2)  \nonumber \\
&& \null 
     - {1\over 4} \Bigl(3 + 2 {x\over y^2} \Bigr) ((X + i\pi)^2 + \pi^2)
     - {2-y^2 \over 4 x^2} ((Y + i\pi)^2 + \pi^2)  
     + {\pi^2\over 6}  \nonumber \\
&& \null
     + {1\over 2} ( 2 x + y^2 ) \biggl[ {1\over y} (X + i\pi) 
                                     + {1\over x} (Y + i\pi) \biggr] 
                 - {1\over 2} \,,
\label{FmpmpSL}
\end{eqnarray}
\begin{eqnarray}
 F^\Sublead_{+--+}(s,t,u)  & = &  
 F^\Sublead_{-+-+}(s,u,t) \,.
\hskip8.5cm
\label{FpmmpSL}
\end{eqnarray}
The variables appearing in the amplitudes are defined in
\eqn{VariableNames}.  

\section{Finite parts of the two-loop $g\gamma \to g\gamma$ amplitudes}
\label{TwoLoopggamggamSection}

The finite remainder for the $g\gamma \to g\gamma$ amplitude is
defined in \eqn{FiniteInfiniteggamggam}.  Its dependence
on quark charges, the renormalization scale $\mu$, $N$ and $\Nf$ 
is given by
\begin{eqnarray}
{\cal M}^{(2) \rm fin}_{g\gamma \to g\gamma} & = &
2 \delta^{ab} \, 
  \, \biggl( \sum_{j = 1}^\Nf Q_j^2 \biggr) \,
S_{\lambda_1 \lambda_2 \lambda_3 \lambda_4} \, \biggl[
   {11 N - 2 \Nf \over 6} \ln(-\musq/u)
    M^{(1)}_{\lambda_1 \lambda_2 \lambda_3 \lambda_4}
\nonumber \\ && \null \hskip 4.5 cm 
  + N G^\Lead_{\lambda_1 \lambda_2 \lambda_3 \lambda_4} 
  - {1\over N} \, 
      F^\Sublead_{\lambda_1 \lambda_2 \lambda_3 \lambda_4} \biggr] \,.
\label{FiniteRemainderggamggam}
\end{eqnarray}
In this case legs 1,3 are the gluons and legs 2,4 are the photons.
The subleading-color functions 
$F^\Sublead_{\lambda_1 \lambda_2 \lambda_3 \lambda_4}$
are the same as for $gg \to \gamma\gamma$.

The leading-color functions 
$G^\Lead_{\lambda_1 \lambda_2 \lambda_3 \lambda_4}$ are new, 
and are given by
\begin{equation}
 G^\Lead_{++++} =  
 {1\over 2} \,,  
\hskip12.5cm 
\label{GppppL}
\end{equation}
\begin{eqnarray}
G^\Lead_{-+++}  & = &  
     {1\over 8}  \biggl[ 
      ( 2 + 4 x y - 5 x^2 ) ( (X-Y)^2 + \pi^2 ) 
    + \biggl(2 + 4 {y\over x^2} - {5\over x^2} \biggr) 
                        ( (Y + i\pi)^2 + \pi^2 ) \nonumber \\
&& \hskip-0.1cm 
       - 2  \biggl( 1 - {x \over y^2} \biggr)
                   ( (X + i\pi)^2 + \pi^2 )
      + 2  \biggl( 9 - 10 {x \over y^2} \biggr) 
        \biggl( y (X-Y) - { y\over x}  (Y + i \pi) \biggr)  \biggr] \,,
 \nonumber \\ && \null
\label{GmpppL}
\end{eqnarray}
\begin{eqnarray}
 G^\Lead_{+-++}  & = &  
     {1\over 8}  \biggl[ 
      ( 2 + 6 x y - 3 x^2 ) ( (X-Y)^2 + \pi^2 ) 
    + \biggl(2 + 6 {y\over x^2} - {3\over x^2} \biggr) 
                        ( (Y + i\pi)^2 + \pi^2 ) \nonumber \\
&& \hskip-0.1cm 
       - 2  { (1 - x)^2 \over y^2 }
                   ( (X + i\pi)^2 + \pi^2 )
      + 2  \biggl( 9 - 8 {x \over y^2} \biggr) 
        \biggl( y (X-Y) - { y\over x}  (Y + i \pi) \biggr)  \biggr] \,,
 \nonumber \\ && \null
\label{GpmppL} 
\end{eqnarray}
\begin{eqnarray}
 G^\Lead_{--++}  & = &  
 - 2 (x^2+y^2) \biggl[ \li4(-x/y) 
     - {1\over2} (X-Y) ( \li3(-x/y) - \zeta_3 )  
     + {\pi^2\over6} \Bigl( \li2(-x/y) 
\nonumber \\ && \hskip0.3cm
     + {1\over2} (X-Y)^2 \Bigr) 
      + {1\over 48} (X-Y)^4 - i {\pi\over12} (X-Y) ( (X-Y)^2 + \pi^2 ) 
      + {17\over720} \pi^4 \biggr] 
\nonumber \\ && \hskip-0.3cm
   + 2 \Bigl( 3 (x-y)^2  - 2 y^2 \Bigr) \biggl[
     \li4(-x/y) + \li4(-x) + \li4(-y)
\nonumber \\ && \hskip3.6cm
   + (Y + i\pi) \Bigl( \li3(-x/y) - \zeta_3 \Bigr)
   + {\pi^2\over6} \Bigl( \li2(-x/y) - {1\over2} Y^2 \Bigr)
\nonumber \\ && \hskip3.6cm
   + {1\over6} (X-Y) Y^3 + {1\over12} Y^4 - {\pi^4\over24} \biggr]
\nonumber \\ && \hskip-0.3cm
   - {2\over3} \biggl( 8 - {x\over y} + 30 x \biggr) \biggr[ 
      \li3(-y) - \zeta_3 - (Y+i\pi)  \Bigl( \li2(-y) - {\pi^2\over6} \Bigr)
\nonumber \\ && \hskip3.4cm
    - {1\over2} (X-Y) ((Y+i\pi)^2 + \pi^2) 
    - {1\over3} Y  \Bigl( Y^2 + {3\over2} i \pi Y + \pi^2 \Bigr) \biggr]
\nonumber \\ && \hskip-0.3cm
 + {1\over6} \biggr( {4\over y} + 27 + 42 y + 4 y^2 \biggr) \biggl[ 
      \li3(-x/y) - \zeta_3 - (X-Y) \biggl( \li2(-x/y) - {\pi^2\over6} \biggr)
\nonumber \\ &&  \hskip4.7cm
    - i {\pi\over2} ((X-Y)^2 + \pi^2) \biggr]
\nonumber \\ && \hskip-0.3cm
+ {1\over12} ( 3 - 2 y - 12 x y ) (X-Y) ((X-Y)^2 + \pi^2)
\nonumber \\ && \hskip-0.3cm
- {1 \over 3 y} (X-Y) ((Y+i\pi)^2 + \pi^2)
+ 2 ( 1 + 2 y ) \biggl( \zeta_3 + {\pi^2\over6} (Y+i\pi) \biggr)
\nonumber \\ && \hskip-0.3cm
+ {1\over24} \biggl( {1\over y^2} - {24\over y} 
                    + 44 - 8 {x^3 \over y^2} \biggr)   ((X+i\pi)^2+\pi^2)
\nonumber \\ && \hskip-0.3cm
- {1\over24} ( 15 - 14 x - 48 x y ) ((X-Y)^2 + \pi^2)
\nonumber \\ && \hskip-0.3cm
+ {1\over24} \biggl( 8 x + 60 - {24\over x} + {27\over x^2} \biggr)  
                                    ((Y+i\pi)^2 + \pi^2)
+ {4\over9} \pi^2 x
\nonumber \\ && \hskip-0.3cm
+ {1\over12}  { 2 x^2 - 54 x y - 27 \over y } 
        \biggl( X-Y - {1\over x} (Y+i\pi) \biggr) \,,
\label{GmmppL}
\end{eqnarray}
\begin{eqnarray}
 G^\Lead_{-+-+}  & = &  
  \biggl( 1 - 2 {x\over y^2} \biggr) \biggl[ 
    4 \li4(-y/x) + 4 \li4(-y) 
  + (3 X - 2 Y + i\pi) \li3(-y/x) 
\nonumber \\ && \hskip2.1cm
  - (X + 2 Y + 3 i\pi) \li3(-y) 
  + ((X-Y)^2 + \pi^2)  \li2(-y/x) 
\nonumber \\ && \hskip2.1cm
  + ((Y+i\pi)^2 + \pi^2) \li2(-y)
  + {1\over8} X^2 (X - 2 Y)^2 
\nonumber \\ && \hskip2.1cm
  - i {\pi\over6} X  \Bigl( (X+i\pi)^2 - 3 X Y \Bigr)  \biggr]
\nonumber \\ && \hskip-0.3cm
 - {1\over2}  \biggl( 1 + 6 {x\over y^2} \biggr) \biggl[ 
      \li3(-x) - \zeta_3 
    - (X+i\pi) \biggl( \li2(-x) - {\pi^2\over6} \biggr)
\nonumber \\ && \hskip2.5cm
    - {1\over6} X (X^2 + 4 \pi^2) 
    + {1\over2} (X-2 Y-i\pi) ((X+i\pi)^2 + \pi^2) \biggr]
\nonumber \\ && \hskip-0.3cm
 - {1\over12}  \biggl( 5 - 2 {x\over y} \biggr)   
                      (X+i\pi) ((X+i\pi)^2 + 3 \pi^2)
 + (X-Y)  ((X+i\pi)^2 + \pi^2) 
\nonumber \\ && \hskip-0.3cm
 + \pi^2 (X+i\pi)
 + {1\over8} \biggl( 14 { x-1 \over y } - 8 y + 9 y^2 \biggr)  
                    ((X-Y)^2 + \pi^2)
\nonumber \\ && \hskip-0.3cm
+ {1\over8} \biggl( 14 {1-x \over y} - 8 {y\over x} 
                   + 9 {y^2 \over x^2} \biggr)  ((Y+i\pi)^2 + \pi^2)
\nonumber \\ && \hskip-0.3cm
+ {1\over8} \biggl( 38 {x\over y^2} - 13 \biggr) ( (X+i\pi)^2 + \pi^2 ) 
\nonumber \\ && \hskip-0.3cm
 - {\pi^2\over6}
 - {9\over4} \biggl[ \biggl( y + 2 {x\over y} \biggr) (X-Y) 
                   - \biggl( {y\over x} + {2\over y} \biggr) (Y+i\pi) \biggr]
 + {1\over2} \,,
\label{GmpmpL}
\end{eqnarray}
\begin{eqnarray}
 G^\Lead_{+--+}  & = &  
  2 { 1+y^2 \over x^2 } \biggl[ 
     \li4(-y) - \zeta_4 
   - {1\over2} (Y+i\pi) ( \li3(-y) - \zeta_3 )
\nonumber \\ && \hskip0.8cm
   + {\pi^2\over6} \biggl( \li2(-y) - {\pi^2\over6} - {1\over2} Y^2 \biggr) 
   - {1\over48} Y^4 
   - {1\over24} (Y+i\pi)^2 ((Y+i\pi)^2 + \pi^2) \biggr]
\nonumber \\ && \hskip-0.3cm
 - 2 { 3 (1-y)^2 - 2 y^2 \over x^2 } \biggl[
\nonumber \\ && \hskip2.5cm
     \li4(-x/y) + \li4(-x) + \li4(-y)
   + (X-Y) ( \li3(-y) - \zeta_3 )
\nonumber \\ && \hskip2.5cm
   + {\pi^2\over6} \biggl( \li2(-y) - X Y - {1\over2} Y^2 \biggr)
   + {1\over4} X^2 Y^2 - {1\over6} X Y^3 - {7\over180} \pi^4 \biggr]
\nonumber \\ && \hskip-0.3cm
 - {2\over3} \biggl( 8 - {1\over y} + {30\over x} \biggr) \biggl[ 
      \li3(-x/y) - \zeta_3 
    - (X-Y) \Bigl( \li2(-x/y) - {\pi^2\over6} \Bigr)
\nonumber \\ && \hskip3.1cm
    + {1\over2} Y ((X-Y)^2 + \pi^2) \biggr]
\nonumber \\ && \hskip-0.3cm
+ {1\over6} \biggl( 4 {x\over y} + 27 
                  + 42 {y\over x} + 4 {y^2\over x^2} \biggr) \biggl[
      \li3(-y) - \zeta_3 
    - (Y+i\pi)  \Bigl( \li2(-y) - {\pi^2\over6} \Bigr)
\nonumber \\ && \hskip5.1cm
    - {1\over3} Y \Bigl( Y^2 + {3\over2} i \pi Y + \pi^2 \Bigr) \biggr]
\nonumber \\ && \hskip-0.3cm
- {1\over12} \biggl( 3 - 2 {y\over x} - 12 {y\over x^2} \biggr)  
                       (Y+i\pi)   ((Y+i\pi)^2 + \pi^2)
\nonumber \\ && \hskip-0.3cm
+ {x \over 3 y} (Y+i\pi) ((X-Y)^2 + \pi^2)
+ 2  \biggl( 1 + 2 {y\over x} \biggr)   
           \biggl( \zeta_3 - {\pi^2\over6} (X-Y) \biggr)
\nonumber \\ && \hskip-0.3cm
+ {1\over24}  \biggl( {x^2\over y^2} - 24 {x\over y} 
                    + 44 - {8 \over x y^2} \biggr)  
                            ( (X+i\pi)^2 + \pi^2 )
\nonumber \\ && \hskip-0.3cm
- {1\over24} \biggl( 15 - {14\over x} - 48 {y\over x^2} \biggr) 
                            ((Y+i\pi)^2 + \pi^2) 
\nonumber \\ && \hskip-0.3cm
+ {1\over24} \biggl( {8\over x} + 60 - 24 x + 27 x^2 \biggr)   
                            ( (X-Y)^2 + \pi^2 )
+ {4\over9}   {\pi^2\over x}
\nonumber \\ && \hskip-0.3cm
+ { 2 - 54 y - 27 x^2 \over 12 y }   
             \biggl( - {1\over x} (Y+i\pi) + (X-Y) \biggr) \,,
\label{GpmmpL}
\end{eqnarray}
where again the variables appearing in the amplitudes are defined in
\eqn{VariableNames}.

In comparison with the one-loop amplitudes for $g\gamma \to g\gamma$,
the two-loop amplitudes are more singular in the forward limit $u \to 0$ 
by one power of $s/u$, due to the exchange of a pair of gluons in the 
$u$-channel.  The power-enhanced terms are only found in the leading 
color functions $G^\Lead$, and only in the helicity configurations
for which the helicity of the incoming gluon does not flip as it scatters
(in the conventional helicity labeling, not the all-outgoing one used 
here).  The helicity of the photon may or may not flip in such terms.  
Explicitly, the power-enhanced terms are
\begin{eqnarray}
G^\Lead_{++++}  & \sim & 0 \,, \nonumber \\
G^\Lead_{-+++}  & \sim & 2 i \pi {s\over u} \,, \nonumber \\
G^\Lead_{+-++}  & \sim & 0 \,, \nonumber \\
G^\Lead_{--++}  & \sim & -{4\over9} (\pi^2+3) i \pi {s\over u} \,,
\nonumber \\
G^\Lead_{-+-+}  & \sim & 0 \,, \nonumber \\
G^\Lead_{+--+}  & \sim & -{4\over9} (\pi^2+3) i \pi {s\over u} \,. 
\label{GLsmallu}
\end{eqnarray}
These terms represent the beginning of the Reggeization of the 
$g\gamma \to g\gamma$ amplitude in perturbation theory; the large
logs of $\ln(-s/u)$ that will arise at subsequent orders in $\alpha_s$
could be resummed using BFKL techniques~\cite{BFKL}.

\section{Checks on results}
\label{ChecksSection}
 
We performed a number of consistency checks on the amplitudes to
ensure their reliability:

\begin{enumerate}
\item As a check of gauge invariance, we verified that the amplitudes
vanish when a gluon or photon polarization vector is replaced with a
longitudinal one.  

\item The agreement of our explicitly computed infrared divergences with
the expected form provides a stringent check on the
amplitudes.  Since the integrals generally contain both divergent and finite
terms, this also provides an indirect verification that the 
leading color finite remainders have been assembled correctly.

\item Using Supersymmetry Ward Identities~\cite{SWI}, we evaluated the
identical helicity case~\cite{TwoLoopSUSY} by relating it to the
already known identical helicity four-gluon
amplitudes~\cite{AllPlusTwo}.  Since the integration was done by a
completely different technique, the agreement between the two
independent ways of evaluating this amplitude provides an additional
stringent check of the programs and integration methods used to obtain
the general helicity cases.

\item We compared the results for $gg \to gg$ obtained using
our methods and computer programs~\cite{ggggpaper} to those of
ref.~\cite{GOTYgggg}.
The interference of the two-loop $gg \to gg$ helicity amplitudes with the
tree amplitudes, after summing over all external helicities and colors
and accounting for the different scheme used (HV {\it vs.} CDR), 
is in complete agreement with the calculation using conventional 
dimensional regularization~\cite{GOTYgggg}.
\end{enumerate}


\section{Conclusions}
\label{ConclusionsSection}

In this paper we presented the two-loop matrix elements for two gluons
to scatter into two photons.  Due to the large gluon-gluon luminosity 
in the $x$ range relevant for the LHC Higgs search, 
these two-loop contributions should be competitive in size with the 
existing next-to-leading order corrections to the 
$q\qb \to \gamma\gamma$ subprocess~\cite{TwoPhotonBkgd1}.  
The matrix elements presented here, together with the associated 
real emission contributions, namely the one-loop amplitudes for 
$gg \to \gamma\gamma g$~\cite{FiveGluon,ggGamGamg}, 
will feed into improved estimates of the QCD background to Higgs 
production at the LHC, when the Higgs decays into 
two photons~\cite{HiggsBkgdPaper}.

\acknowledgments
We thank Adrian Ghinculov and Carl Schmidt for useful conversations.



\begin{thebibliography}{99}

\bibitem{HiggsRadCorr}
G.~Degrassi,
hep-ph/0102137; \\
J.~Erler,
hep-ph/0102143.

\bibitem{SusyHiggs}
M.~Carena, H.E.~Haber, S.~Heinemeyer, W.~Hollik, C.E.~Wagner and G.~Weiglein,
Nucl.\ Phys.\ B {\bf 580}, 29 (2000)
[hep-ph/0001002]; \\
J.R.~Espinosa and R.~Zhang,
Nucl.\ Phys.\ B {\bf 586}, 3 (2000)
[hep-ph/0003246].

\bibitem{LEPHiggs}
R.~Barate {\it et al.}  [ALEPH Collaboration],
Phys.\ Lett.\ B {\bf 495}, 1 (2000)
[hep-ex/0011045];
P.~Abreu {\it et al.}  [DELPHI Collaboration],
Phys.\ Lett.\ B {\bf 499}, 23 (2001)
[hep-ex/0102036];
M.~Acciarri {\it et al.}  [L3 Collaboration],
Phys.\ Lett.\ B {\bf 508}, 225 (2001)
[hep-ex/0012019].
G.~Abbiendi {\it et al.}  [OPAL Collaboration],
Phys.\ Lett.\ B {\bf 499}, 38 (2001)
[hep-ex/0101014].

\bibitem{Higgsgammagamma}
J.~R.~Ellis, M.~K.~Gaillard and D.~V.~Nanopoulos,
Nucl.\ Phys.\ B {\bf 106}, 292 (1976); \\
M.~A.~Shifman, A.~I.~Vainshtein, M.~B.~Voloshin and V.~I.~Zakharov,
Sov.\ J.\ Nucl.\ Phys.\  {\bf 30}, 711 (1979)
[Yad.\ Fiz.\  {\bf 30}, 1368 (1979)]; \\
J.F.~Gunion, P.~Kalyniak, M.~Soldate and P.~Galison,
Phys.\ Rev.\ D {\bf 34}, 101 (1986); \\
J.F.~Gunion, G.L.~Kane and J.~Wudka,
Nucl.\ Phys.\ B {\bf 299}, 231 (1988).

\bibitem{HiggsBkgdgammagamma}
R.K.~Ellis, I.~Hinchliffe, M.~Soldate and J.J.~van der Bij,
Nucl.\ Phys.\ B {\bf 297}, 221 (1988).

\bibitem{TwoPhotonBkgd1}
P.~Aurenche, A.~Douiri, R.~Baier, M.~Fontannaz and D.~Schiff,
Z.\ Phys.\ C {\bf 29}, 459 (1985); \\
B.~Bailey, J.F.~Owens and J.~Ohnemus,
Phys.\ Rev.\ D {\bf 46}, 2018 (1992); \\
B.~Bailey and J.F.~Owens,
Phys.\ Rev.\ D {\bf 47}, 2735 (1993); \\
B.~Bailey and D.~Graudenz,
Phys.\ Rev.\ D {\bf 49}, 1486 (1994)
[hep-ph/9307368]; \\
C.~Balazs, E.L.~Berger, S.~Mrenna and C.P.~Yuan,
Phys.\ Rev.\ D {\bf 57}, 6934 (1998)
[hep-ph/9712471]; \\
C.~Balazs and C.P.~Yuan,
Phys.\ Rev.\ D {\bf 59}, 114007 (1999)
[Erratum-ibid.\ D {\bf 63}, 059902 (1999)]
[hep-ph/9810319].
T.~Binoth, J.P.~Guillet, E.~Pilon and M.~Werlen,
Eur.\ Phys.\ J.\ C {\bf 16}, 311 (2000)
[hep-ph/9911340]; 
Phys.\ Rev.\ D {\bf 63}, 114016 (2001)
[hep-ph/0012191]; \\
T.~Binoth,
hep-ph/0005194.

\bibitem{AmetllerDicusWillenbrock}
L.~Ametller, E.~Gava, N.~Paver and D.~Treleani,
Phys.\ Rev.\ D {\bf 32}, 1699 (1985); \\
D.A.~Dicus and S.S.~Willenbrock,
Phys.\ Rev.\ D {\bf 37}, 1801 (1988).

\bibitem{HiggsBkgdPaper}
Z. Bern, L. Dixon and C. Schmidt, in preparation.

\bibitem{BRY}
Z.~Bern, J.S.~Rozowsky and B.~Yan,
Phys.\ Lett.\ B {\bf 401}, 273 (1997)
[hep-ph/9702424];\\
%
Z.~Bern, L.~Dixon, D.C.~Dunbar, M.~Perelstein and J.S.~Rozowsky,
Nucl.\ Phys.\ B {\bf 530}, 401 (1998)
[hep-th/9802162].

\bibitem{AllPlusTwo}
Z.~Bern, L.~Dixon and D.A.~Kosower,
JHEP {\bf 0001}, 027 (2000)
[hep-ph/0001001].

\bibitem{BhabhaTwoLoop}
Z.~Bern, L.~Dixon and A.~Ghinculov,
Phys.\ Rev.\ D {\bf 63}, 053007 (2001)
[hep-ph/0010075].

\bibitem{GOTY2to2}
C.~Anastasiou, E.W.~Glover, C.~Oleari and M.E.~Tejeda-Yeomans,
Nucl.\ Phys.\ B {\bf 601}, 318 (2001)
[hep-ph/0010212];\\
%
C.~Anastasiou, E.W.~Glover, C.~Oleari and M.E.~Tejeda-Yeomans,
Nucl.\ Phys.\ B {\bf 601}, 341 (2001)
[hep-ph/0011094];\\
%
C.~Anastasiou, E.W.~Glover, C.~Oleari and M.E.~Tejeda-Yeomans,
Nucl.\ Phys.\ B {\bf 605}, 486 (2001)
[hep-ph/0101304].

\bibitem{GOTYgggg}
E.W.~Glover, C.~Oleari and M.E.~Tejeda-Yeomans,
Nucl.\ Phys.\ B {\bf 605}, 467 (2001).

\bibitem{CutBased}
W.L.~van Neerven,
Nucl.\ Phys.\ {\bf B268}, 453 (1986); \\
Z.~Bern, L.~Dixon, D.C.~Dunbar and D.A.~Kosower,
Nucl.\ Phys.\ {\bf B425}, 217 (1994)
[hep-ph/9403226]; \\
Z.~Bern, L.~Dixon, D.C.~Dunbar and D.A.~Kosower,
Nucl.\ Phys.\ B {\bf 435}, 59 (1995)
[hep-ph/9409265];\\
Z.~Bern, L.~Dixon and D.A.~Kosower,
Ann.\ Rev.\ Nucl.\ Part.\ Sci.\ {\bf 46}, 109 (1996)
[hep-ph/9602280].

\bibitem{PBScalar}
V.A.~Smirnov,
Phys.\ Lett.\  {\bf B460}, 397 (1999)
[hep-ph/9905323].

\bibitem{NPBScalar}
J.B.~Tausk,
Phys.\ Lett.\  {\bf B469}, 225 (1999)
[hep-ph/9909506].

\bibitem{PBReduction}
V.A.~Smirnov and O.L.~Veretin,
Nucl.\ Phys.\  {\bf B566}, 469 (2000)
[hep-ph/9907385].

\bibitem{NPBReduction}
C.~Anastasiou, T.~Gehrmann, C.~Oleari, E.~Remiddi and J.B.~Tausk,
Nucl.\ Phys.\  {\bf B580}, 577 (2000)
[hep-ph/0003261].

\bibitem{GRReduction}
T.~Gehrmann and E.~Remiddi,
Nucl.\ Phys.\  {\bf B580}, 485 (2000)
[hep-ph/9912329].

\bibitem{IntegralsAGO}
C.~Anastasiou, E.W.N.~Glover and C.~Oleari,
Nucl.\ Phys.\  {\bf B565}, 445 (2000)
[hep-ph/9907523]; \\
C.~Anastasiou, E.W.N.~Glover and C.~Oleari,
Nucl.\ Phys.\  {\bf B575}, 416 (2000),
err. ibid.\  {\bf B585}, 763 (2000)
[hep-ph/9912251].

\bibitem{Catani}
S.~Catani,
Phys.\ Lett.\  {\bf B427}, 161 (1998) [hep-ph/9802439].

\bibitem{GieleGlover}
W.T.~Giele and E.W.~Glover,
Phys.\ Rev.\ D {\bf 46}, 1980 (1992).

\bibitem{KST}
Z.~Kunszt, A.~Signer and Z.~Tr\'ocs\'anyi,
Nucl.\ Phys.\ B {\bf 420}, 550 (1994)
[hep-ph/9401294].

\bibitem{FiveGluon}
Z.~Bern, L.~Dixon and D.A.~Kosower,
Phys.\ Rev.\ Lett.\  {\bf 70}, 2677 (1993)
[hep-ph/9302280].

\bibitem{ggGamGamg}
D.~de Florian and Z.~Kunszt,
Phys.\ Lett.\ B {\bf 460}, 184 (1999)
[hep-ph/9905283];\\
%
C.~Balazs, P.~Nadolsky, C.~Schmidt and C.P.~Yuan,
Phys.\ Lett.\ B {\bf 489}, 157 (2000)
[hep-ph/9905551].

\bibitem{OneLoopIR}
See e.g.,
W.T.~Giele, E.W.~Glover and D.A.~Kosower,
Nucl.\ Phys.\ B {\bf 403}, 633 (1993)
[hep-ph/9302225];\\
%
S.~Frixione, Z.~Kunszt and A.~Signer,
Nucl.\ Phys.\ B {\bf 467}, 399 (1996)
[hep-ph/9512328];\\
%
S.~Catani and M.H.~Seymour,
Phys.\ Lett.\ B {\bf 378}, 287 (1996)
[hep-ph/9602277];
%
Nucl.\ Phys.\ B {\bf 485}, 291 (1997)
[Erratum-ibid.\ B {\bf 510}, 503 (1997)]
[hep-ph/9605323].

\bibitem{HV}
G.~'t Hooft and M.~Veltman,
Nucl.\ Phys.\ {\bf B44}, 189 (1972).

\bibitem{SchemeConvert}
Z.~Bern and D.A.~Kosower,
Nucl.\ Phys.\ {\bf B379}, 451 (1992).
%
Z.~Kunszt, A.~Signer and Z.~Tr\'ocs\'anyi, 
Nucl.\ Phys.\ {\bf B411}, 397 (1994)
[hep-ph/9305239].
%
S.~Catani, M.H.~Seymour and Z.~Tr\'ocs\'anyi,
Phys.\ Rev.\ {\bf D55}, 6819 (1997)
[hep-ph/9610553].

\bibitem{BernMorgan}
Z.~Bern and A.G.~Morgan,
Nucl.\ Phys.\ B {\bf 467}, 479 (1996)
[hep-ph/9511336].

\bibitem{SpinorHelicity}
F.A.~Berends, R.~Kleiss, P.~De Causmaecker, R.~Gastmans and T.T.~Wu,
Phys.\ Lett.\ B {\bf 103}, 124 (1981);\\
%
P.~De Causmaecker, R.~Gastmans, W.~Troost and T.T.~Wu,
Phys.\ Lett.\ B {\bf 105}, 215 (1981);\\
%
Z.~Xu, D.~Zhang and L.~Chang,
Nucl.\ Phys.\ B {\bf 291}, 392 (1987).

\bibitem{MPReview}
M.L.~Mangano and S.J.~Parke,
Phys.\ Rept.\ {\bf 200}, 301 (1991);\\
L.~Dixon,
in {\it Proceedings of Theoretical Advanced Study Institute in
Elementary Particle Physics (TASI 95)}, ed.\ D.E.\ Soper
[hep-ph/9601359].

\bibitem{KosowerHelicity}
D.A.~Kosower,
Phys.\ Lett.\ B {\bf 254}, 439 (1991).

\bibitem{DimShift}
Z.~Bern, L.~Dixon and D.A.~Kosower,
Phys.\ Lett.\ B {\bf 302}, 299 (1993)
[Erratum-ibid.\ B {\bf 318}, 649 (1993)]
[hep-ph/9212308];\\
%
Z.~Bern, L.~Dixon and D.A.~Kosower,
Nucl.\ Phys.\ B {\bf 412}, 751 (1994)
[hep-ph/9306240].

\bibitem{TwoLoopSUSY}
Z.~Bern, A.~DeFreitas, L.~Dixon and H.L.~Wong,
to appear.

\bibitem{ggggpaper}
Z.~Bern, A. De Freitas, L.~Dixon and A.~Ghinculov, in preparation.

\bibitem{NielsenRef}
See e.g. K.S.~K\"olbig,
SIAM J.\ Math.\ Anal.\  {\bf 17}, 1232 (1986).

\bibitem{Lewin}
L.~Lewin, {\it Dilogarithms and Associated Functions} (Macdonald, 1958).

\bibitem{NielsenIds}
K.S.~K\"olbig, J.A.~Mignaco and E.~Remiddi, B.I.T. {\bf 10}, 38 (1970).

\bibitem{PhotonPaper}
Z. Bern, A. De Freitas, L. Dixon, A. Ghinculov and H.L. Wong, 
preprint SLAC--PUB--8974, UCLA/01/TEP/18, hep-ph/0109079

\bibitem{BFKL}
E.~A.~Kuraev, L.~N.~Lipatov and V.~S.~Fadin,
Sov.\ Phys.\ JETP {\bf 44}, 443 (1976)
[Zh.\ Eksp.\ Teor.\ Fiz.\  {\bf 71}, 840 (1976)]; \\ 
%
E.A.~Kuraev, L.N.~Lipatov and V.S.~Fadin,
Sov.\ Phys.\ JETP {\bf 45}, 199 (1977)
[Zh.\ Eksp.\ Teor.\ Fiz.\  {\bf 72}, 377 (1977)]; \\
%
I.I.~Balitsky and L.N.~Lipatov,
Sov.\ J.\ Nucl.\ Phys.\  {\bf 28}, 822 (1978)
[Yad.\ Fiz.\  {\bf 28}, 1597 (1978)].

\bibitem{SWI}
M.T.~Grisaru, H.N.~Pendleton and P.~van Nieuwenhuizen,
Phys.\ Rev.\ D {\bf 15}, 996 (1977); \\
M.T.~Grisaru and H.N.~Pendleton,
Nucl.\ Phys.\ B {\bf 124}, 81 (1977); \\
S.J.~Parke and T.R.~Taylor,
Phys.\ Lett.\ B {\bf 157}, 81 (1985),
err. ibid.\  {\bf 174B}, 465 (1985).

\end{thebibliography}
\end{document}